\documentclass[10pt, journal]{IEEEtran}
\IEEEoverridecommandlockouts
\usepackage{graphicx}
\usepackage{subfig}
\usepackage{amsmath,amsthm,amsfonts,amssymb}
\usepackage{cite}
\usepackage{bm}
\usepackage{bbm}
\usepackage{url}
\usepackage{array}
\usepackage{color}
\usepackage{multirow}
\usepackage{booktabs}
\usepackage[table,xcdraw]{xcolor}
\usepackage{enumerate}
\usepackage{epstopdf}
\usepackage{tabularx}
\usepackage{algorithm}
\usepackage[english]{babel}
\usepackage{algpseudocode}

\begin{document}
\title{\makebox[\linewidth]{\parbox{\dimexpr\textwidth+0cm\relax}{\centering Addressing Out-of-Distribution Challenges in Image Semantic Communication Systems with Multi-modal Large Language Models}}}

\author{
\IEEEauthorblockN{Feifan Zhang$^*$, \IEEEmembership{Student Member, IEEE}, Yuyang Du$^*$, \IEEEmembership{Student Member, IEEE}, Kexin Chen$^*$, \IEEEmembership{Student Member, IEEE}\\Yulin Shao, \IEEEmembership{Member, IEEE}, Soung Chang Liew, \IEEEmembership{Fellow, IEEE}} \vspace{-2em}
\IEEEauthorblockA{\thanks{F. Zhang, Y. Du and S. C. Liew are with the Department of Information Engineering, The Chinese University of Hong Kong, Hong Kong SAR, China. K. Chen is with the Department of Computer Science Engineering, The Chinese University of Hong Kong, Hong Kong SAR, China. Y. Shao is with the State Key Laboratory of Internet of Things for Smart City, University of Macau, Macau SAR, China.}}
\IEEEauthorblockA{\thanks{The work was partially supported by the Shen Zhen-Hong Kong-Macao technical program (Type C) under Grant No. SGDX20230821094359004.}} 
\IEEEauthorblockA{\thanks{S. C. Liew is the corresponding author (e-mail: soung@ie.cuhk.edu.hk).}} 
\IEEEauthorblockA{\thanks{$^*$Authors contribute equally to this work.}} 
}

\maketitle

\begin{abstract}
Semantic communication is a promising technology for next-generation wireless networks. However, the out-of-distribution (OOD) problem, where a pre-trained machine learning (ML) model is applied to unseen tasks that are outside the distribution of its training data, may compromise the integrity of semantic compression. This paper explores the use of multi-modal large language models (MLLMs) to address the OOD issue in image semantic communication. We propose a novel ``Plan A - Plan B" framework that leverages the broad knowledge and strong generalization ability of an MLLM to assist a conventional ML model when the latter encounters an OOD input in the semantic encoding process. Furthermore, we propose a Bayesian optimization scheme that reshapes the probability distribution of the MLLM's inference process based on the contextual information of the image. The optimization scheme significantly enhances the MLLM's performance in semantic compression by 1) filtering out irrelevant vocabulary in the original MLLM output; and 2) using contextual similarities between prospective answers of the MLLM and the background information as prior knowledge to modify the MLLM's probability distribution during inference. Further, at the receiver side of the communication system, we put forth a ``generate-criticize" framework that utilizes the cooperation of multiple MLLMs to enhance the reliability of image reconstruction.
\end{abstract}


\section{Introduction}\label{sec-I}
From the first generation (1G) to the fifth generation (5G), advanced communication techniques have significantly improved the transmission rate of wireless networks. The upcoming 6G network is designed to support even higher transmission rates \cite{Ref000,dang2020should,Ref043d,Ref039,Ref004,Ref040}, facilitating diverse communication-reliant applications, such as augmented reality (AR), virtual reality (VR), and autonomous driving. However, bit-based transmission technologies are currently approaching the Shannon Capacity \cite{arikan2009channel}. This motivates researchers to develop goal-oriented communication systems beyond bit transmission.

Semantic communication transcends the traditional bit-wise transmission by prioritizing the meaning and intent conveyed through information exchange. It effectively compresses data by filtering out irrelevant information while preserving semantic fidelity, significantly reducing the bandwidth requirement for information transmission \cite{xie2021deep,Ref041,gunduz2022beyond}. As such, semantic communication is expected to become a key technology in next-generation wireless communication systems.

A semantic communication system consists of a semantic transmitter and a semantic receiver. At the transmitter side, the transmitter first performs semantic encoding, in which the sender extracts semantic features from the source data and compresses the intended message in a way that causes no semantic information loss. Subsequently, channel coding is applied to the semantic features to generate signals suitable for physical-channel transmission. The two encoding modules can be implemented separately, as in a conventional communication system, or they can be implemented as a joint source-channel coding (JSCC) integrated system. At the receiver side, reverse operations are carried out to reconstruct the semantic contents \cite{Ref010,Ref042,Ref011}.

Many existing works realized semantic communication systems with machine learning (ML) models, such as convolutional neural network (CNNs) or transformers \cite{Ref012,Ref013,Ref014,Ref015}. In \cite{Ref012}, for example, the authors proposed a Generative Adversarial Network (GAN)-based model for image semantic coding and decoding. In \cite{Ref013}, the authors put forth a generative semantic communication framework based on scene graphs, which uses a graph neural network (GNN) for semantic encoding at the transmitter and a diffusion model for decoding at the receiver. In \cite{Ref014}, the authors introduced a JSCC framework for wireless image transmission. It employs convolutional neural networks to directly map image pixel values to complex transmission signals, demonstrating robust performance in environments with low signal-to-noise ratios (SNRs). A later work \cite{Ref015} proposed a receiver-feedback JSCC scheme that utilizes the transformer's self-attention mechanism to boost the discriminative representation of semantic features.

\begin{figure}[htbp]
\centering
\subfloat[]{\label{fig1a} \includegraphics[width=0.23\textwidth]{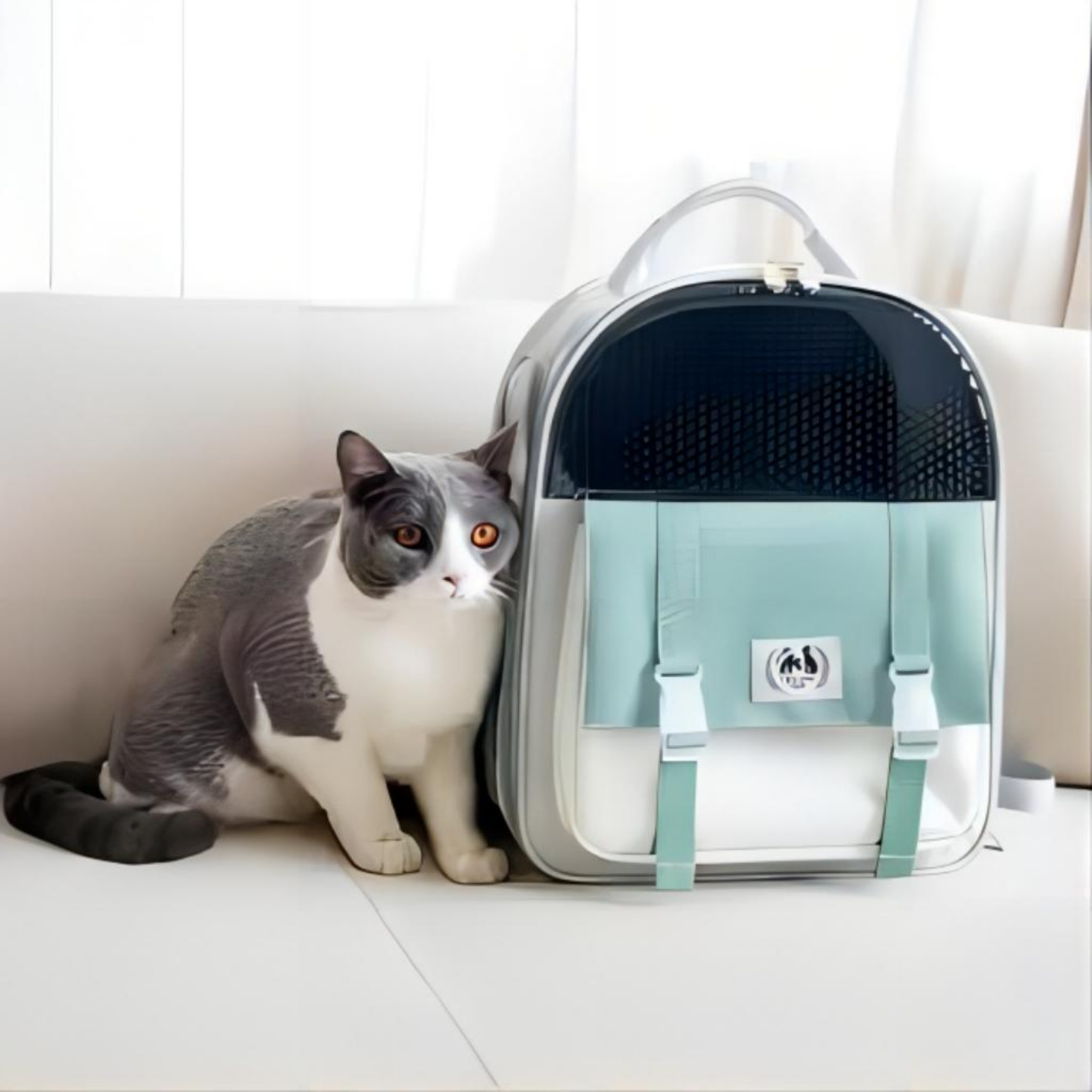}}
\subfloat[]{\label{fig1b} \includegraphics[width=0.23\textwidth]{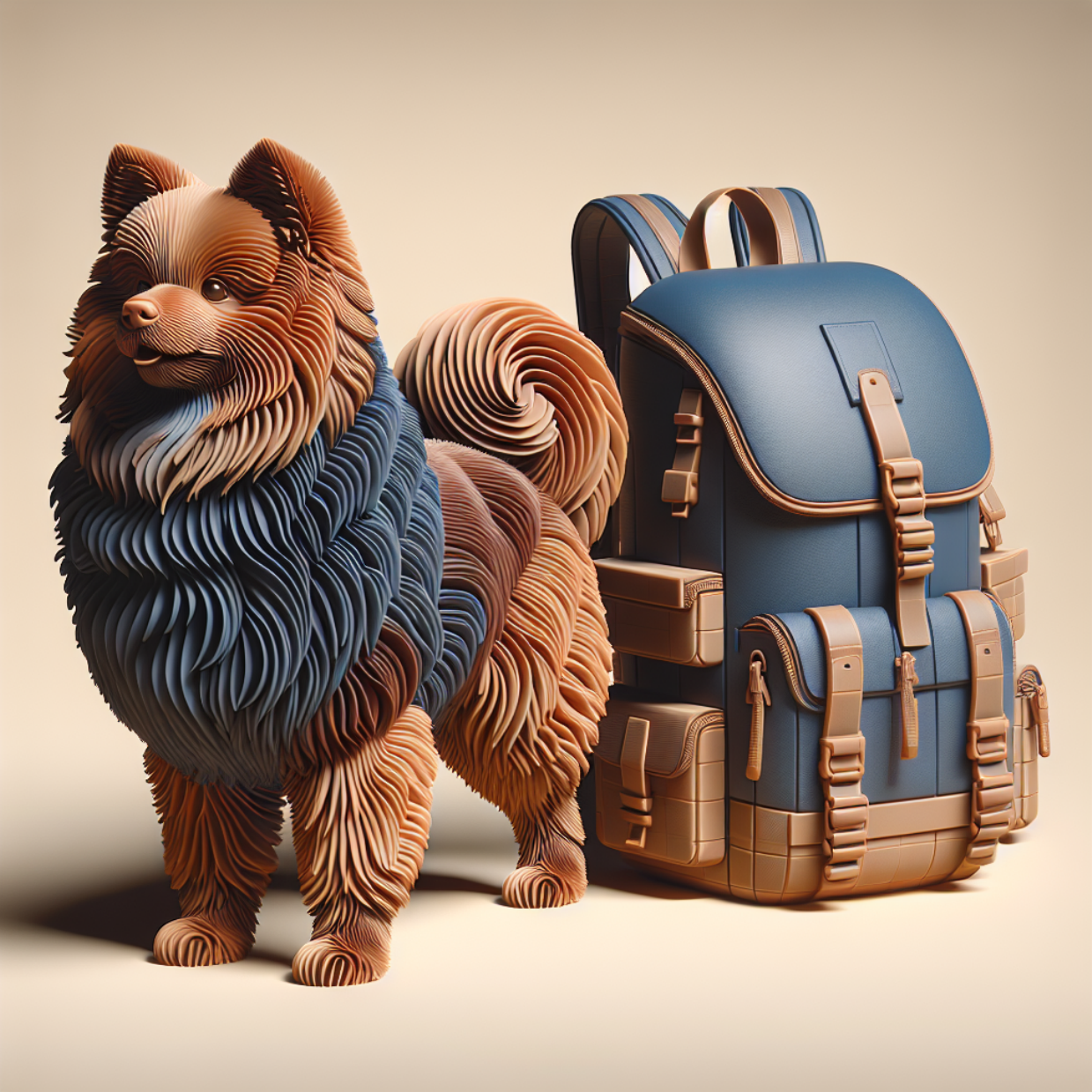}}
\captionsetup{font={small}}
\caption{(a) The original image. (b) A reconstructed image with the OOD problem.}
\label{fig:figure1}
\end{figure}

A major issue not addressed in the above papers is the out-of-distribution (OOD) problem prevalent in conventional ML models \cite{Ref016}. An ML model in a semantic communication system is typically trained on a specific dataset assumed to be representative of the data distribution for the target task. The model learns to extract important semantic features based on the training data. However, when the pre-trained model is applied to a new, unseen task that is out of the distribution of the training data, the ML model may fail to correctly extract the true semantic information. Fig. \ref{fig:figure1} provides an example in which the training dataset of the semantic encoder does not contain knowledge about cats due to its limited scale. Although the semantic encoder can identify the backpack in the image, it has no idea about the object besides the backpack. The semantic encoder might misclassify the cat as another type of animal it knows from its training data, such as a dog, thus introducing semantic distortion at the transmitter. Subsequently, the receiver reconstructs the image with the distorted semantic information and obtains a significantly different image in Fig. \ref{fig:figure1}b.

The dynamic and ever-changing nature of the real world inevitably introduces OOD data to realistic semantic communication systems, as novel situations, events, and entities continuously emerge. Therefore, building a practical semantic encoder with a strong ability to handle OOD data is crucial. This paper explores the use of multi-modal large language models (MLLMs), which converts a given input image into a textual description, to address the challenge. A conventional ML model's training dataset may focus on a specific domain, and the model can perform quite well given tasks within its knowledge distribution. On the other hand, MLLMs are trained on extremely large-scale and general datasets \cite{Ref038a,Ref038b,Ref038c}. Compared with a conventional ML model's expertise in a domain-specific task, MLLMs have a broad knowledge base that can be applied to various fields and tasks, although they may not perform as well as an expert ML model in domain-specific tasks for which the ML model is specially trained. This paper demonstrates that the vast knowledge and generalization ability of a MLLM can be a valuable augmentation to a conventional ML model to overcome the OOD problem encountered in semantic communication. Given the general knowledge of the MLLM and the domain specific expertise of the conventional ML model, the rest of paper refers to these two models as ``general AI model" and ``expert AI model", respectively.


This paper focuses mainly on the semantic compression process happening within the semantic encoder at the transmitter side. For system-building and demonstration purposes, we also deploy off-the-shelf generative AI models for image reconstruction at the receiver side. Furthermore, we investigate the reliability performance of the communication system in the face of strong channel noise. Our contributions are summarized as follows:

\textbf{A ``Plan A -- Plan B" Framework}: At the transmitter side, we put forth a novel ``Plan A -- Plan B" framework that overcome the OOD problem in the semantic coding process with the help of MLLMs. For non-OOD cases, the expert AI expresses confidence in its judgment. Thus, we take the classification result of the expert model as ``Plan A". On the other hand, for OOD cases, the expert AI expresses extremely low confidence in its judgment (which, in essence, is just a random guess). Therefore, when the confidence of the expert AI's judgment is below a pre-defined threshold, we turn to the general AI and treat its classification result as an alternative ``Plan B". When OOD occurs, our scheme, with the general AI's augmentation, significantly outperforms the conventional expert-AI-empowered semantic communication system in semantic extraction. Our experimental results indicate a 13.3\% classification accuracy improvement on a comprehensive dataset with 8.2\% OOD and 91.8\% non-OOD cases. For pure OOD cases, the classification accuracy of the semantic encoding process increases from close to 0 to 71.9\%.

\textbf{A Tailored Output Layer Using Bayesian Optimization}: The object identification within the encoding process expects answers that precisely describe the image with a set of nouns. However, the original output of a MLLM may contain many irrelevant words, including verbs, adjectives, articles, or even greetings like ``hello" and ``hi", which is useless for describing an image. Our recent work \cite{Ref017} developed a novel cross-entropy transformation (CET) scheme that narrows the vocabulary of the output for precise object identification purposes. With CET, we map the complex self-attention matrix extracted from the last fully connected layer of the transformer network, which considers the full vocabulary list, into a probability distribution among target classes.

Building upon the CET scheme, this paper optimizes the extracted probability distribution via Bayesian analysis with the help of the image's contextual information. When the expert model delegates the identification task to the MLLM, a straightforward but naive approach for the general AI is to choose the class with the highest probability as the identification result. However, the target object may sometimes be difficult to identify due to irrelevant details. For example, consider a plate with the logo of a cartoon bear. As far as semantic communication is concerned, the major information to be conveyed should be the plate, but the general AI might get confused between the classes ``bear" and ``plate" if the naive approach were applied. A remedial approach is to leverage background information of the target object. Specifically, a realistic image typically contains multiple objects. Even if an expert AI cannot identify an object with high confidence because of OOD, it might still be able to clearly identify other objects within the figure. The collection of these clearly identifiable objects can serve as useful contextual information to the MLLM. Returning to the example, if we tell the MLLM that the background of the image includes a dining table, cakes, teacups, knives, and forks, then the MLLM might be induced to conclude that the correct classification is ``plate" with very high confidence, as a plate looks more harmonious and coherent with the kitchen environment than a bear does.

We realize the above intuition with Bayes theorem. Given the MLLM's probability distribution, we first calculate the contextual similarity between each prospective class and the background information within the image. Then, we reshape the MLLM's probability distribution with Bayesian probability, using the normalized contextual similarity of each class to the background information as the known knowledge, and the MLLM's original probability distribution as a priori probability, and finally obtain the corresponding a posterior probability distribution. In complex cases where the highest probability of the MLLM's prediction does not significantly outweigh the second or third highest probabilities, the reshaped probability distribution might give these runner-up classes another chance to be considered. This could be helpful when the model's confidence in its top prediction is not substantially greater than its confidence in the next best alternatives, as illustrated in the above ``bear \& plate" example.

The CET scheme, which generates the probability distribution on a narrowed vocabulary list, and the Bayesian optimization, which reshapes the probability distribution, together form a novel output layer tailored to our semantic compression task with MLLM.

\textbf{A ``Generate-Criticize" Framework for Image Reconstruction:} At the receiver side, we propose a ``generate-criticize" framework that utilizes multi-MLLM cooperation to enhance the reliability of image reconstruction. Specifically, the receiver side deploys a text-to-image MLLM and an image-to-text MLLM, which are referred to as the image generation model and the criticizing model, respectively. As their names suggest, the first MLLM generates an image according to the text description of the desired image (which is recovered from the wireless signals received), while the second MLLM compares the generated image with the text description and points out the flaws of the generated image so that the first MLLM can revise its output accordingly.

The generate-criticize framework is inspired by an interesting observation during our experiments. We found that existing text-to-image MLLMs like Dall-E-3 \cite{Ref018} and Stable Diffusion \cite{Ref019} are poor at handling numbers (similar observations have also been reported in \cite{Ref020, Ref021}). However, such mistakes can be discovered by image-to-text MLLMs like GPT-4V \cite{Ref022}. Therefore, to mitigate the semantic distortion caused by the image generation MLLM's poor handling of numbers (e.g., number of similar objects in an image), we deploy an image-to-text MLLM to challenge the generated image. Then in the next iteration, the image generation MLLM can revise the image according to the critic's feedback. Our experimental results indicate that the ``generate-criticize" framework significantly enhances the quality of image generation, boosting the probability of correct image generation by around 20\% with up-to-four-round iterations.

\section{System Model and Problem Formulation}\label{sec-II}
\begin{figure}[htbp]
\centering
\subfloat[]{\label{fig2a} \includegraphics[width=0.23\textwidth]{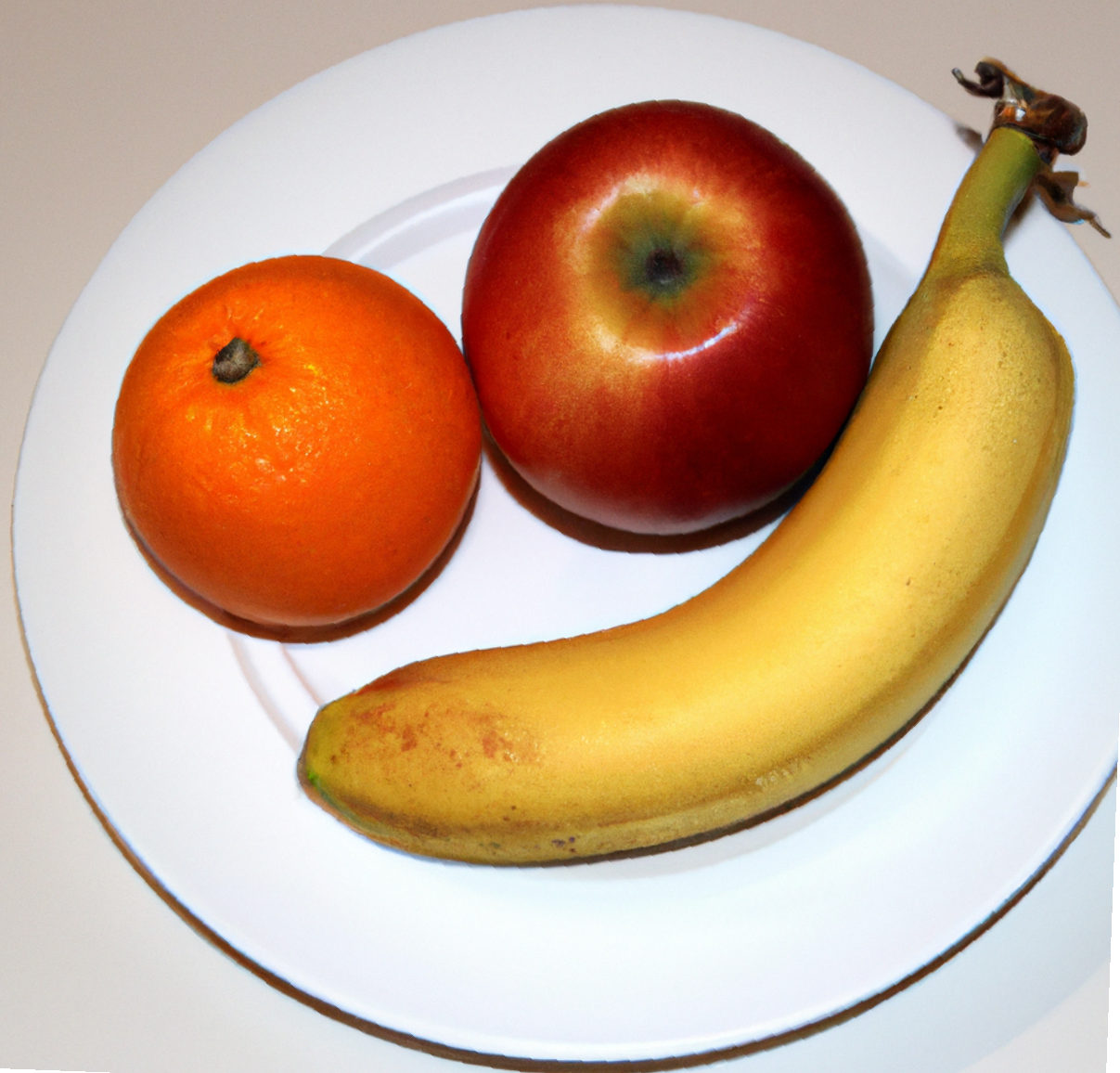}}
\subfloat[]{\label{fig2b} \includegraphics[width=0.23\textwidth]{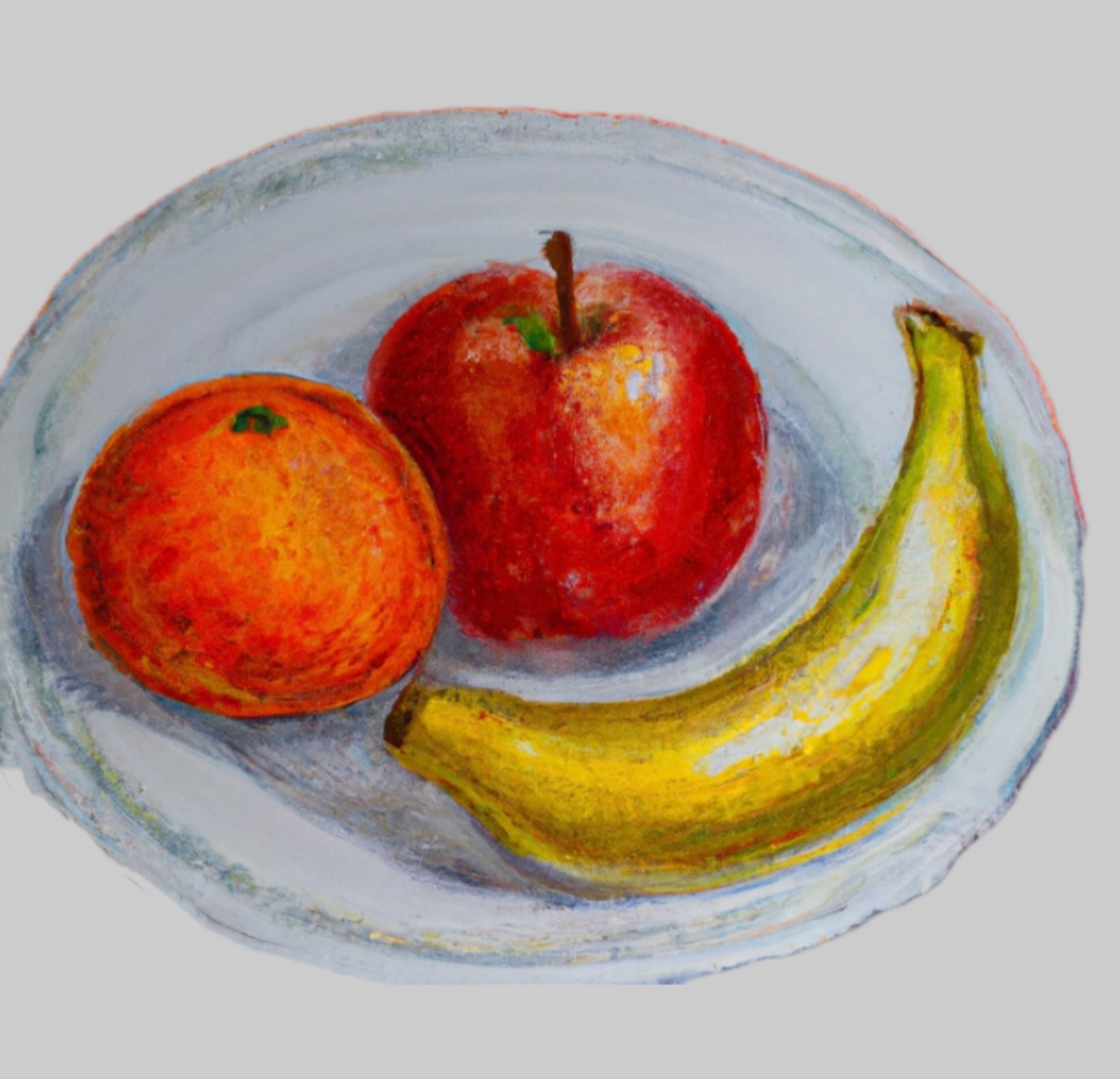}}
\captionsetup{font={small}}
\caption{(a) The photo of a plate and some fruits. (b) The oil painting of the same fruit plate.}
\label{fig:figure2}
\end{figure}

Many previous investigations in semantic communication focus on pixel-wise image fidelity \cite{Ref012,Ref013,Ref014,Ref015,Ref024}. This paper, on the other hand, focuses on accurate element-wise semantic representation of the image. When interpreting an image, humans seldom focus on the RGB values of the pixels at the outset. Instead, they are more likely to first concentrate on identifying the objects or elements in the figure and interpreting what the figure tries to convey by analyzing the juxtaposition of these objects. In other words, when reading an image, humans focus more on the element-wise semantic information rather than the pixel-wise image information. Therefore, as long as the element-wise semantic meaning of a figure can be successfully recovered at the receiver, the transmission can be considered as largely correct, regardless of specific pixel values. For example, Fig. \ref{fig2a} presents a plate with some fruits, while Fig. \ref{fig2b} renders the same scenario as an oil painting. Although the two figures are completely different from the perspective of a pixel-based analysis, they convey essentially the same meaning from a semantic perspective: a banana, an apple, and an orange are placed on a white plate.

In this section, we will first introduce our system framework and define the associated notations. Then, we present the loss function assumed in this paper that considers the integrity of semantic information from two angles: 1) the correct recovery of the transmitted element-wise information, and 2) the perceptual similarity between the transmitted image and the recovered image.

We define the element-wise semantic information of an image by $\bm{S}=[S_1,S_2,\ldots,S_n]$, where $n$ is the total number of elements contained in the image, and $S_i$ represents the $i^{th}$ element among them. For element $S_i$, it could be the background of the image or an object in the image.

\begin{figure*}[htbp]
  \centering
  \includegraphics[width=0.98\textwidth]{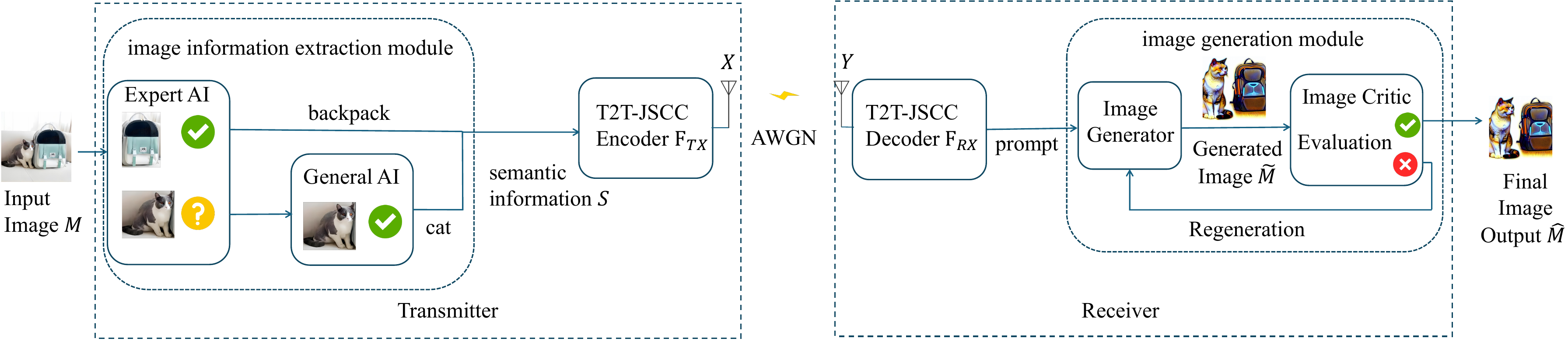}\\
  \captionsetup{font={small}}
  \caption{ The framework of the proposed image semantic transmission system.}\label{fig:figure3}
  \vspace{-1.5mm}
\end{figure*}

Fig. \ref{fig:figure3} presents the overall framework of our semantic communication system. At the beginning of the transmission process, semantic encoding is conducted to extract the semantic information from image $M$. The encoding process first attempts to identify the elements within the image using a conventional ML model trained with domain-specific data. For element $S_i$, let us denote the expert model's confidence in its judgment by $C_i$. As discussed in the introduction, the expert model cannot precisely identify an element that never appears in its training data, and it may have a relatively low $C_i$ in an OOD case. If $C_i$ is below a pre-defined threshold $\rho$, we extract a sub-image that solely contains $S_i$ by cropping the whole image (to avoid confusing the general AI in the identification task of other objects) and provide the sub-image to the general AI for further processing.

This paper trains YOLOv8 \cite{Ref026}, a lightweight open-source CNN model, as the expert AI, and InstructBLIP \cite{Ref027}, a large open-source visual-language MLLM, as the general AI. We refer readers to Section III for details on 1) how we train the expert AI with a domain-specific dataset, 2) how we adapt the general AI with our Bayesian-optimized output layer, and 3) how the image is jointly processed by the expert AI and the general AI.

After extracting the element-wise semantic information $\bm{S}=[S_1,S_2,\ldots,S_n]$ with the joint effort of the expert AI and the general AI, the transmitter inputs $\bm{S}$ to a text-to-text joint source-channel coding (T2T-JSCC) encoder.\footnote{After the element-wise semantic compression, the textual description of the image can be transmitted via a conventional communication system with source coding (e.g., Huffman coding), channel coding (e.g., convolution coding), and modulation. Alternatively, we could apply the T2T-JSCC module for the same purpose. This paper chooses the later for system building.} As in \cite{Ref028, Ref029, Ref030}, the T2T-JSCC encoder combines the conventional source coding, channel coding, and modulation into a single encoder and parameterizes the module by $F_{TX}(\cdot)$. After the semantic element $S_i$ is encoded by the T2T-JSCC encoder, its corresponding output $X_i$ becomes a normalized representation in a $d$-dimensional vector space, where $d$ is a hyperparameter of the T2T-JSCC encoder. We write $X_i$ as:
\begin{equation}
X_i = F_{TX}(S_i) \in \mathbb{C}^d \text{ .}
\end{equation}

With the above definition, the output of the T2T-JSCC encoder can be written as:
\begin{equation}
\bm{X} = F_{TX}(\bm{S}) = [F_{TX}(S_1), F_{TX}(S_2), \ldots, F_{TX}(S_n)] \text{ .}
\end{equation}

For more information about the implementation and analysis of the T2T-JSCC encoder, we refer readers to Section II and Section III of \cite{Ref028} for details.

Vector $\bm{X}$ is the signal to be transmitted over the noisy channel. For $X_i$, its corresponding channel output $Y_i$ can be written as:
\begin{equation}
Y_i = H(X_i) = X_i + N_i \text{ ,}
\end{equation}
where $N_i$ denotes the Gaussian noise over the transmitted signal $X_i$.

At the receiver side, the system applies a T2T-JSCC decoder to decode the received signal $\bm{Y}=[Y_1,Y_2,\ldots,Y_n]$. The T2T-JSCC decoder is symmetrical to the encoder. We parameterize the T2T-JSCC decoder by $F_{RX}(\cdot)$ and write the decoded message as:
\begin{equation}
\bm{\hat{S}} = F_{RX}(Y) = [\hat{S}_1, \hat{S}_2, \ldots, \hat{S}_n] \text{ .}
\end{equation}

After obtaining $\bm{\hat{S}}$ through T2T-JSCC decoding, the receiver composes the text information into a prompt that describes the image. The prompt is then given to the image reconstruction module.

The image reconstruction module aims to build an image $\hat{M}$ that is consistent with the semantic information described in the given prompt. The image reconstruction module contains two MLLMs playing complementary roles. The first MLLM is realized with DALL-E-3 \cite{Ref018}. It takes text as input and generates images according to the given description. The second MLLM is realized with GPT-4V \cite{Ref022}. It analyzes the generated image and determines whether the image is consistent with the description in the prompt. If the generated image perfectly reflects the prompt, we take the image as the final output. Otherwise, the critic MLLM modifies the original prompt by adding suggestions on how the image should be revised to better reflect the semantic information. The new prompt is returned to DALL-E-3 for image generation in the next round. The above iteration continues until the critic MLLM is satisfied with the image (i.e., the image is consistent with the semantic information provided in the prompt) or the number of iterations reaches an upper limit.

We define the loss function for our semantic communication system as
\begin{equation}
L = (1 - \alpha) L_1 \langle \bm{S}, \bm{\hat{S}} \rangle + \alpha L_2 \langle M, \hat{M} \rangle \text{ ,}
\end{equation} \label{loss_function}
where $L_1 \langle \bm{S}, \bm{\hat{S}} \rangle$ denotes the textual dissimilarity between the lists of elements at the transmitter side and the receiver side, falling within [0, 1]; $L_2 \langle M, \hat{M} \rangle$ reflects the visual distance between the original image and the recovered image, falling within [0, 1]; and $1-\alpha$ and $\alpha$, $0<\alpha<1$, denote the weights of $L_1$ and $L_2$, respectively.

It is important to note that typical measurements of textual dissimilarity/similarity must be conducted within a high-dimensional semantic space generated by a natural language processing (NLP) network. Therefore, to calculate $L_1 \langle \bm{S}, \bm{\hat{S}} \rangle$, we need to transform $\bm{S}$ and $\bm{\hat{S}}$ into corresponding vectors within the high-dimensional semantic space. For that purpose, we first concatenate the elements within $\bm{S}$ and $\bm{\hat{S}}$, and then we leverage an NLP embedding network (this paper uses SentenceBERT \cite{Ref031} and Spacy \cite{Ref036} as two examples) to transform the element concatenations into two corresponding vectors within the semantic space of the NLP embedding network. Let us denote the dimension of the semantic space by $G_1$, and denote the two output vectors of the NLP embedding network by $\bm{T}$ and $\bm{\hat{T}}$, respectively.

With $\bm{T}$ and $\bm{\hat{T}}$, we calculate the cosine angle between the two vectors, which is denoted by $CA \langle \bm{T}, \bm{\hat{T}} \rangle$ and could be written as
\begin{equation}
CA \langle \bm{T}, \bm{\hat{T}} \rangle = \frac{\sum_{n=1}^{G_1} T_n \hat{T}_n}{\sqrt{\sum_{n=1}^{G_1} (T_n)^2 \cdot \sum_{n=1}^{G_1} (\hat{T}_n)^2}} \text{ .}
\end{equation}

We note that a large $CA \langle \bm{T}, \bm{\hat{T}} \rangle$ indicates a small angle between $\bm{T}$ and $\bm{\hat{T}}$, corresponding to high similarity between $\bm{S}$ and $\bm{\hat{S}}$. Thus, we define the dissimilarity between $\bm{T}$ and $\bm{\hat{T}}$ as
\begin{equation}
L_1 \langle \bm{S}, \bm{\hat{S}} \rangle = 1 - CA \langle \bm{T}, \bm{\hat{T}} \rangle \text{ .}
\end{equation}

For $L_2 \langle M, \hat{M} \rangle$, the perceptual image loss during the communication process, this paper applies Learned Perceptual Image Patch Similarity (LPIPS) \cite{Ref032} as a measurement of distance\footnote{In \cite{Ref032}, the authors referred to their metric as ``similarity". However, as we can see from equation (1) in \cite{Ref032}, the metric increases when the differences between the two images become more distinct. In other words, LPIPS is, in essense, a \textit{distance} or \textit{dissimilarity} rather than a \textit{similarity}. Although this paper follows the traditional name developed in \cite{Ref032} for consistency, we point out this naming inconsistency to prevent readers' potential confusion.}. Compared with the traditional pixel-wise metrics, LPIPS aims to mimic the perceptual results of human vision, thereby better conforming to the way human judgment measures the similarity between two images. LPIPS utilizes deep neural networks (this paper uses AlexNet \cite{Ref033} as an example) to compute the distance between the transmitted image $M$ and the reconstructed image $\hat{M}$ in the feature space of the network to measure the dissimilarity of the two images. Specifically, $L_2 \langle M, \hat{M} \rangle$ is given by
\begin{equation}
L_{2}\langle M, \hat{M}\rangle=\sum_{l} \frac{1}{H_{l} W_{l}} \sum_{h, w}\left\|w_{l} \odot\left(y_{h w}^{l}-\hat{y}_{h w}^{l}\right)\right\|_{2}^{2} \text{ ,}
\label{LPIPs}
\end{equation}
where $H_l, W_l, C_l$ represent the height, width, and channel dimension of the feature generated by AlexNet at the $l^{th}$ layer; $y_{hw}^l \in \mathbb{R}^{H_l \times W_l \times C_l}$ is the feature that AlexNet generates at the $l^{th}$ layer according to the transmitted image, while $\hat{y}_{hw}^l \in \mathbb{R}^{H_l \times W_l \times C_l}$ is the feature that AlexNet generates at the $l^{th}$ layer according to the reconstructed image; and $\odot$ represents the channel-wise feature multiplication operation. For more explanations on (\ref{LPIPs}) given above, we refer readers to Section III of \cite{Ref032} for more details.

\section{Transmitter Design}\label{sec-III}
This section focuses on the design of our semantic transmitter. We will present implementation details of our novel ``Plan A – Plan B" framework and Bayesian-optimized output layer in subsection \ref{sec-III-A} and subsection \ref{sec-III-B}, respectively. Then, in subsection \ref{sec-III-C}, we provide a brief introduction to how we adapt the T2T-JSCC encoder reported in \cite{Ref028} for our task.
\subsection{The Plan A – Plan B Framework} \label{sec-III-A}
We leverage the YOLOv8 project for the realization of the expert AI. The original YOLOv8 model reported in \cite{Ref026} was built upon the COCO dataset \cite{Ref037}, which contains over 330,000 images and 80 object categories. For concept-proving purposes, this paper reorganizes the training dataset of YOLOv8 to manually construct an OOD scenario. Specifically, among the 80 classes within the COCO dataset, there are ten labels related to animals, such as ``cat", ``bird", and ``elephant". We removed all images containing animal label(s) from the training dataset but kept the testing dataset unchanged. Consequently, the model learns nothing about animals from the training data but faces animal images in the testing dataset. For the realization of the general AI, we use InstructBLIP's pre-trained model weights reported in \cite{Ref027} and adapt its output layer accordingly (see the discussion in Section \ref{sec-III-B}).

We now introduce the implementation of the Plan A – Plan B framework. We first applied Plan A, the expert AI model, for object identification of the input image. Assume that there are $n$ objects in the image. For object $i$, the expert model draws a rectangle to frame the object and generates the corresponding recognition result $S_{i,A}$, where the subscript $A$ denotes that this is the recognition result of Plan A. Note that $S_{i,A}$ should be one of the labels within the revised COCO dataset, which has 70 classes only.

Let us denote the expert model's confidence in $S_{i,A}$ by $C_i$. If $C_i$ is below a pre-defined threshold $\rho$, we treat the identification result as unreliable. For such cases, we rely on Plan B, the general AI, for a reliable semantic encoding process. We crop the image within the rectangle frame to obtain a sub-image that contains the object solely. We then give the sub-image to InstructBLIP, which performs an object classification task with a simple prompt (i.e., ``What is in the figure?") and generates $S_{i,B}$. Note that $S_{i,B}$ is one of the labels within the original COCO dataset (with 80 classes that include animal-related ones). We refer readers to subsection \ref{sec-III-B} for how the classification task is conducted within the MLLM. 

After InstructBLIP corrects the recognition results of all low-confidence objects, we finally obtain the semantic information vector $\textbf{S} =[S_1,S_2,…,S_n ]$, in which 
\begin{equation}
S_{i}=\left\{\begin{array}{ll}
S_{i, A}, & C_{i} \geq \rho \\
S_{i, B}, & C_{i}<\rho
\end{array}\right. \text{ .}
\end{equation}

\subsection{The Bayesian Optimized Output Layer} \label{sec-III-B}
This subsection considers the case where the expert AI is not confident in its identification of element $i$ (i.e., $C_i<\rho$) and shifts the identification task to the general AI. We discuss how the general AI obtains the probability distribution for its identification task and revises the probability distribution with a Bayesian method.

Let us start with a quick introduction to the general AI model. InstructBLIP is an open-source MLLM with vision and language ability. There are three components in InstructBLIP: 1) an image encoder that deals with the image input, 2) a text-in-text-out language model to manage the output, and 3) an image-text transformer that bridges the input and output modules. Thanks to the modular architectural design, InstructBLIP is highly flexible, and we can quickly adapt a wide range of text-to-text language models for implementation. Without loss of generality, our implementation utilizes FlanT5 \cite{Ref034}, an instruction-tuned model based on Transformer T5 \cite{Ref035}, as the text-in-text-out model.

We next introduce the CET scheme. The CET scheme has been reported in detail in our previous work \cite{Ref017}, and this paper only gives a quick introduction at a conceptual level. We refer interested readers to Section II-A of \cite{Ref017} for illustrations and implementation details.

For the general AI, given the sub-image of object $i$, let us denote the self-attention matrix at the last fully connected layer of FlanT5 by $\bm{e}$. The size of $\bm{e}$ is $P\times P$, where $P$ is the cardinality of InstructBLIP's full vocabulary list after tokenization (let us denote the full vocabulary list by $\bm{l^F}$). Element $e_{k1,k2}$ within $\bm{e}$ is the attention weight between word $k_1$ and word $k_2$ within $\bm{l^F}$. Let us denote the number of possible classes in our classification task by $M$, and denote the list of these classification labels (also after tokenization) by $\bm{l^S}$, in which the superscript $S$ indicates that the list is essentially a subset of the full vocabulary list $\bm{l^F}$. Given $\bm{l^S}$, the CET scheme extracts the corresponding $M$ rows from the original self-attention matrix $\bm{e}$. Let us denote the new $M\times P$ matrix by $\bm{e^E}$, where the superscript $E$ notes that the matrix is a smaller one after extracting the $M$ rows. Meanwhile, with the tokenized label list $\bm{l^S}$ (of size $M\times 1$) and the tokenized full vocabulary list $\bm{l^F}$ (of size $P\times 1$), we conduct one-hot encoding and obtain an $M\times P$ matrix $\bm{b}$, in which
\begin{equation}
b_{j, k}=\left\{\begin{array}{ll}
1, & l_{j}^{S}=l_{k}^{F} \\
0, & l_{j}^{S} \neq l_{k}^{F}
\end{array}\right. \text{ .}
\end{equation}

With $\bm{e^R}$ and $\bm{b}$, we calculate the cross-entropy loss between the two $M\times P$ matrices by,
\begin{equation}
L_{j}=-\sum_{x=1}^{P} e_{j, x}^{R} \cdot b_{j, x} \text{ .}
\end{equation}

In the resulting $M\times 1$ vector $\bm{L}$, element $L_j$ represents the expected loss of using classification label $j$ as the identification result. Note that $L_j$ is negatively related to the probability of label $j$ being selected. Therefore, a reciprocal process is needed for transforming $L_j$ to an intermediate variable $D_j$ that is positively related to the probability of label $j$ being selected. This paper follows the realization in \cite{Ref017} and lets $D_j=-L_j$. This is followed by the normalization process. Eventually, at the end of the CET scheme, we obtain a normalized $M\times 1$ probability vector $\bm{P}$, in which element $P_j$ denotes the probability of label $j$ being selected.

\begin{figure}[htbp]
\centering
\subfloat[]{\label{fig4a} \includegraphics[width=0.24\textwidth]{Figures/Figure4/fig4a.pdf}}
\subfloat[]{\label{fig4b} \includegraphics[width=0.24\textwidth]{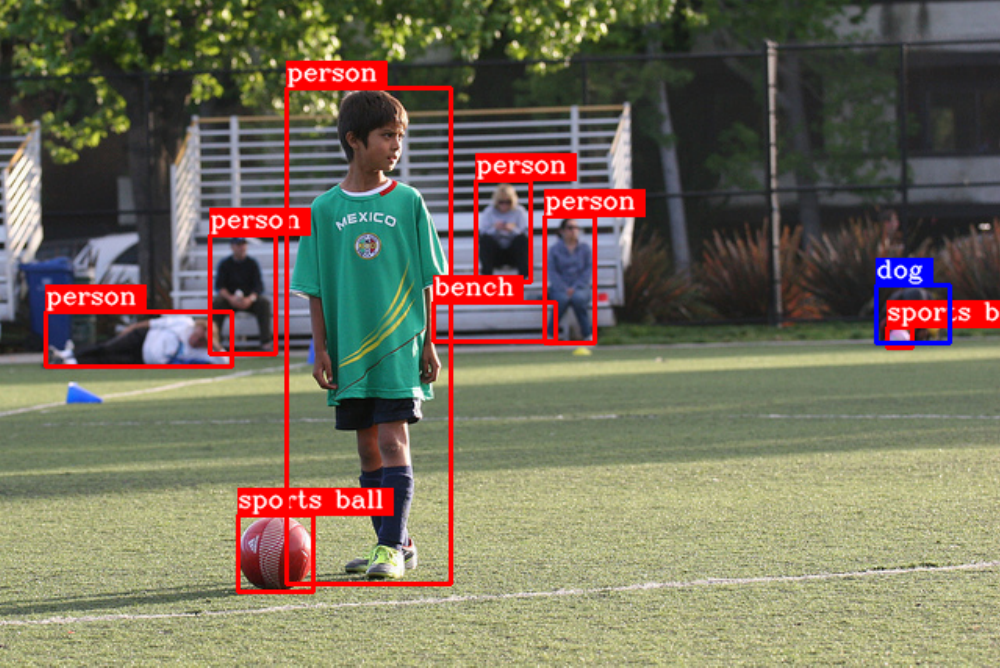}}
\vspace{-3.0mm}
\quad
\captionsetup{justification=centering}
\subfloat[]{\label{fig4c} \includegraphics[width=0.15\textwidth]{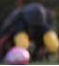}}
\subfloat[]{\label{fig4d} \includegraphics[width=0.16\textwidth]{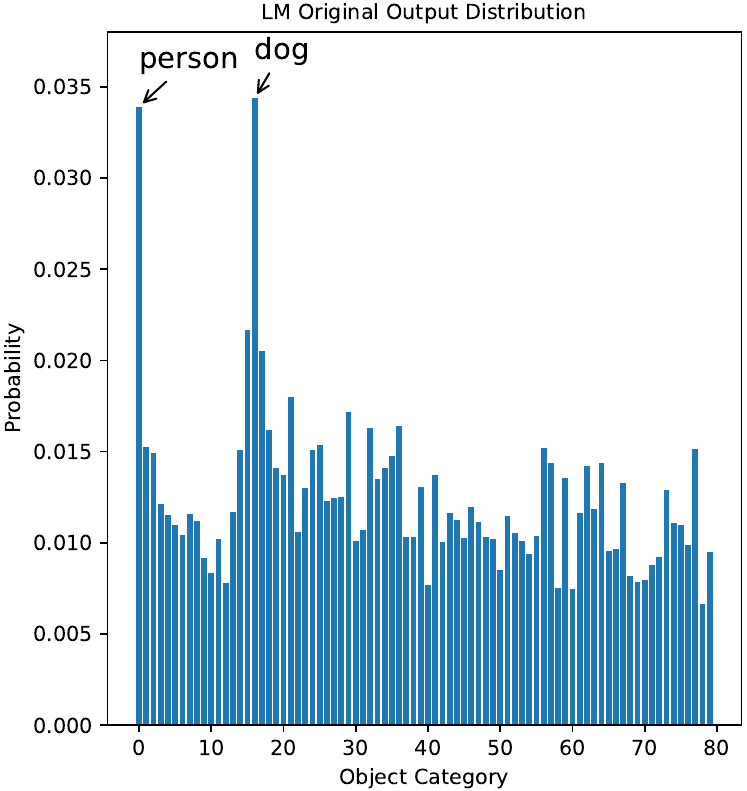}}
\subfloat[]{\label{fig4e} \includegraphics[width=0.165\textwidth]{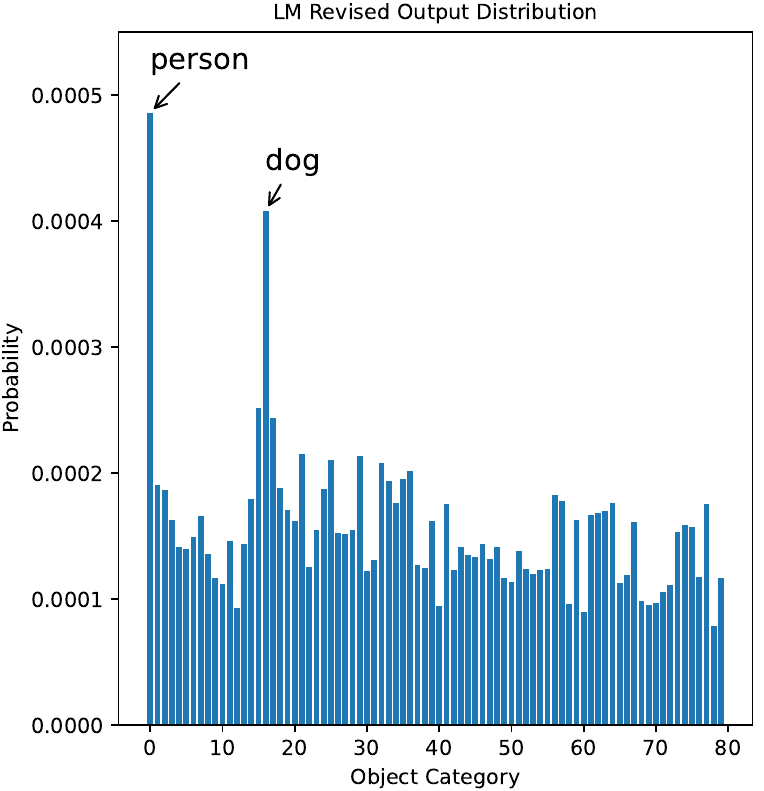}}
\vspace{0.0mm}
\captionsetup{font={small}}
\caption{A specific example of our Bayesian optimization scheme.}
\label{fig:figure4}
\end{figure}

After the CET scheme, we now move on to our Bayesian optimization scheme with the illustration of Fig. \ref{fig:figure4}. Fig. \ref{fig:figure4}a shows an image with a person playing and watching sports balls on a field, in which the blurry objects in the background may be difficult for the semantic extraction process, even with the Plan A – Plan B framework.\footnote{The Plan A – Plan B framework not only helps with the OOD scenarios (as we have argued in the introduction), but is also very effective in handling fuzzy images (as we show in Fig. \ref{fig:figure4} and discussions related to it). Our later experiment in Section V validates that point.}  Fig. \ref{fig:figure4}b illustrates the identification results after the expert AI and the general AI (without our Bayesian optimization method described below) collaboratively extract the figure's semantics. In the figure, identified objects are marked with rectangles along with the corresponding identification results. For the reader's easier observation, we mark the misidentified object with a blue rectangle and mark the rest of the correctly identified ones with red rectangles. There is one misidentified case in Fig. \ref{fig:figure4}b. It shows a person kneeling on the ground (see Fig. \ref{fig:figure4}c for a zoom-in image). However, due to the fuzziness and unclear object features, both the expert AI and the general AI fail to clearly identify the person. The probability distribution of the general AI presented in Fig. \ref{fig:figure4}d indicates that the general AI does get confused between ``person" and ``dog".

Let us refer to the set of elements that can be successfully identified by the expert AI solely as the contextual information and denote it by $\bm{I}$ expressed as
\begin{equation}
\bm{I} = {S_{i,A}, \forall i \text{ s.t. } C_i \geq \rho} \text{ .}
\end{equation}
In the investigated case presented in Fig. \ref{fig:figure4}b, for example, the contextual information $\bm{I}$ is: {``person", ``person", ``person", ``person", ``person", ``sports ball", ``sports ball", ``bench"}.

The aim of our Bayesian optimization scheme is to reshape the general AI's original probability distribution (i.e., the one presented in Fig. \ref{fig:figure4}d) to a more reasonable one (i.e., the one presented in Fig. \ref{fig:figure4}e) with the additional information provided by the contextual information within the image. For that purpose, we need to calculate the similarity between $I'$ and each label. Let us denote the contextual similarity between $I'$ and $l_j^S$ by $CS\langle I', l_j^S \rangle$. Similar to the calculation of textual dissimilarity discussed in Section \ref{sec-II}, we concatenate all elements within $\bm{I}$ and denote the concatenation by $I'$. After that, we apply an NLP embedding network to map $I'$ and $l_j^S$ into a $G_2$-dimensional semantic space. Let us denote the embedding vectors of $I'$ and $l_j^S$ by $\bm{T^I}$ and $\bm{T^j}$, wherein the superscripts $I$ and $j$ indicate the input of contextual information and label $j$, respectively. Then we calculate $CA\langle \bm{T^I}, \bm{T^j} \rangle$, the cosine angle between $\bm{T^I}$ and $\bm{T^j}$, as,
\begin{equation}
CA\langle \bm{T^I}, \bm{T^j} \rangle = \frac{\sum_{n=1}^{G_2} T_n^I T_n^j}{\sqrt{\sum_{n=1}^{G_2} (T_n^I)^2 \cdot \sum_{n=1}^{G_2} (T_n^j)^2}} \text{ .}
\end{equation}

It is important to note that different embedding networks may result in semantic spaces of different dimensions. Some semantic spaces are sensitive to the similarity between input texts, while others are not (depending on the training material and implementation details of each embedding network). In other words, even with the same $I'$ and $l_j^S$, cosine angle $CA\langle \bm{T^I}, \bm{T^j} \rangle$ may change when different embedding networks are applied. Yet, we note that the cosine angle always falls within $[0,1]$, no matter which network is applied. We introduce an exponential index $\tau$ to $CA\langle \bm{T^I}, \bm{T^j} \rangle$ and define $CS\langle I', l_j^S \rangle$ as
\begin{equation}
CS\langle I', l_j^S \rangle = [CA\langle \bm{T^I}, \bm{T^j} \rangle]^\tau \text{ .}
\end{equation}

With the above definition, $CS\langle I', l_j^S \rangle$ falls within $[0,1]$, and the difference between embedding networks (in terms of their sensitivity to contextual similarity) can be well compensated by the appropriate setting of index $\tau$. Further, once the embedding network to be applied in our system is decided, the value of $\tau$ could be adjusted with the training dataset without worrying about the OOD problem (given that the optimal setting of $\tau$ is related only to the embedding network's sensitivity to similarity).

With $CS\langle I', l_j^S \rangle$ and the general AI's original probability distribution $\bm{P}$ (the one directly extracted with the CET scheme), we now calculate the classification probability of label $j$ as,
\begin{equation}
P_j' = \frac{CS\langle I', l_j^S \rangle}{\sum_{n=1}^M P_j \cdot CS\langle I', l_n^S \rangle} \cdot P_j \text{ ,}
\end{equation}
where $P_j'$ is the revised probability of label $j$ being selected as the classification output.

With the revised probability distribution $\bm{P'}$, we choose the label with the highest probability as the revised classification result of the general AI model.
\subsection{The T2T-JSCC encoder} \label{sec-III-C}
The T2T-JSCC encoder transforms semantic information vector $\bm{S}$ into an $N \times d$ complex vector $\bm{X}$ that is ready for transmission over the noisy wireless channel. The processing inside the T2T-JSCC encoder includes source coding, channel coding, and digital modulation.

Since we do not need to transmit a very long sentence as in \cite{Ref028}, we re-developed the T2T-JSCC encoder with the following adjustments. To begin with, we follow the baseline method in \cite{Ref028} and utilize differentiable cross-entropy as the loss function. Furthermore, given that the simplified task does not require a very complex semantic space, we reduce the dimension of the output signal by setting a smaller hyperparameter $d$ (specifically, from $128$ to $50$). Additionally, during the training process, we use object labels from the COCO dataset as new training materials to retrain the model.

With the above modifications, our new T2T-JSCC encoder is more lightweight than the original one reported in \cite{Ref028}. However, it is stable and reliable enough to meet our requirements and demonstrates strong anti-noise performance (as proven in the following experimental section).

\section{Receiver Design}\label{sec-IV}
This section focuses on the design of our semantic receiver. We propose a ``generate-criticize" image reconstruction framework for more reliable reconstruction based on semantic information recovered from the T2T-JSCC decoder.

\begin{figure}[htbp]
\centering
\subfloat[]{\label{fig5a} \includegraphics[width=0.23\textwidth]{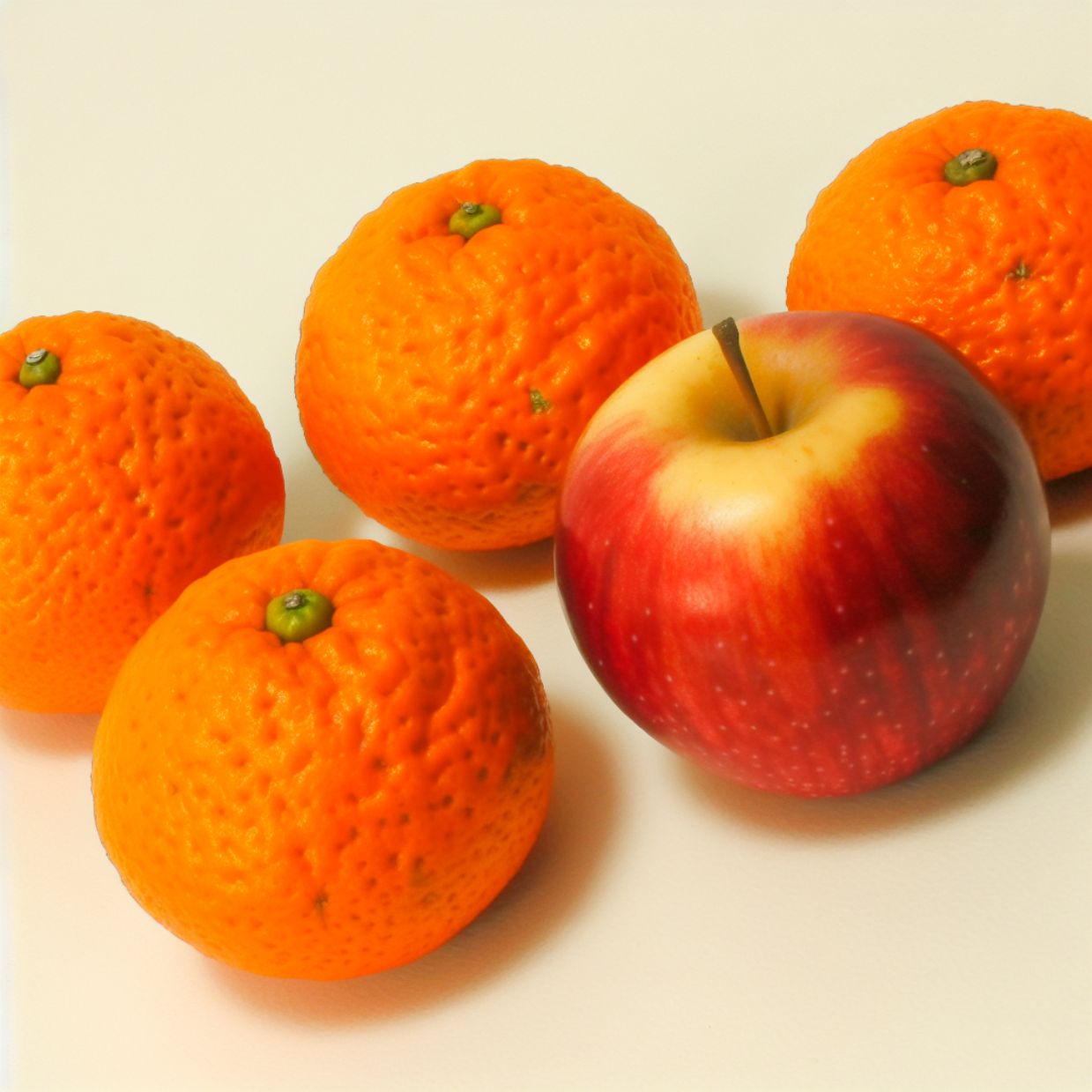}}
\subfloat[]{\label{fig5b} \includegraphics[width=0.23\textwidth]{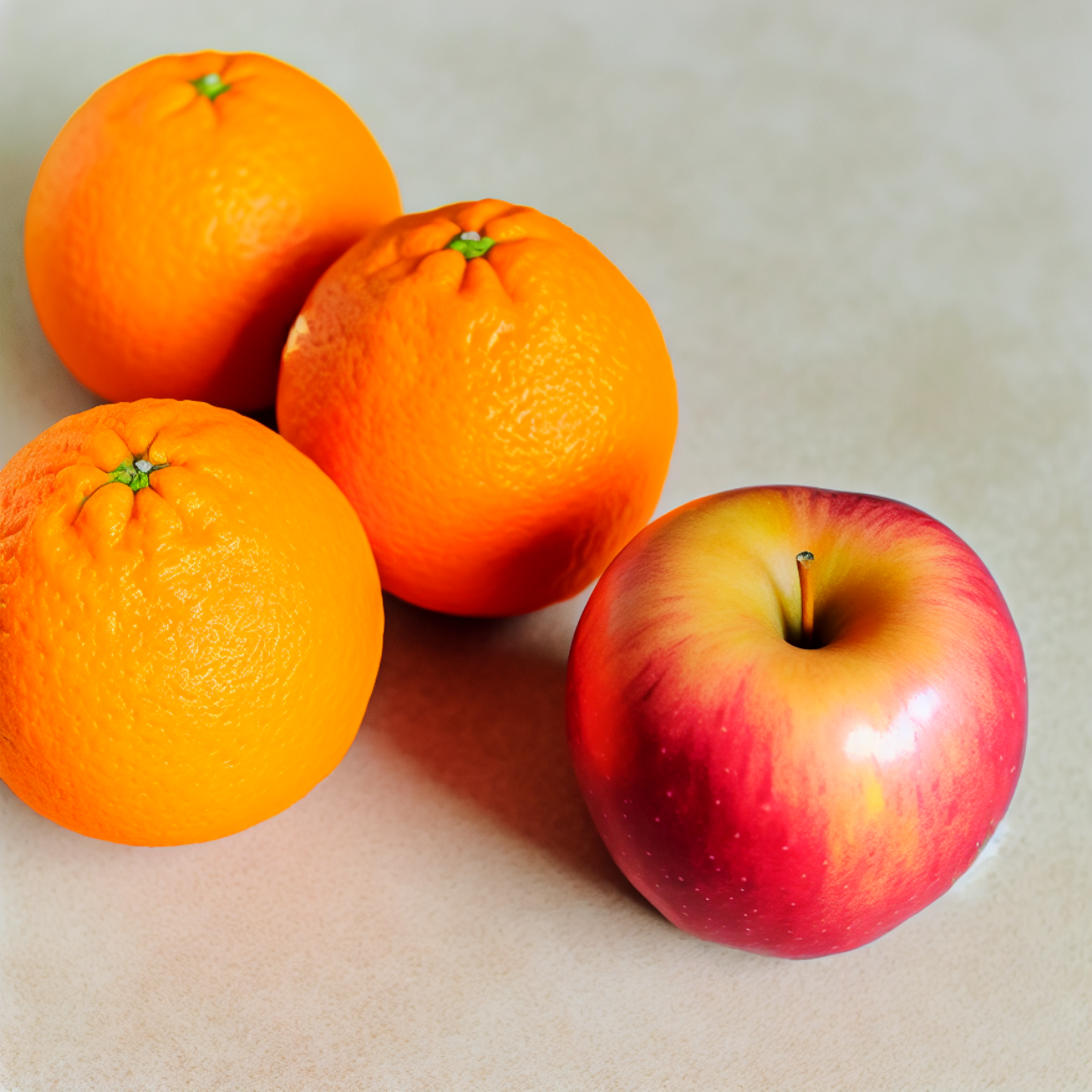}}
\captionsetup{font={small}}
\caption{An example of the ``generate-criticize" image reconstruction framework.}
\label{fig:figure5}
\end{figure}

At the receiver side, the recovered semantic information is a vector containing the labels of all identified objects within the transmitted image. For example, let's say the transmitted image contains three oranges and one apple, and the successfully recovered semantic vector is: $\bm{\hat{S}} = [\text{orange}, \text{apple}, \text{orange}, \text{orange}]$. We then asked DALL-E-3 to reconstruct this image based on this vector and obtained the image shown in Fig. \ref{fig:figure5}a. However, the generated image exhibits inconsistencies: the image contains one apple and four oranges, which does not align with the original object counts. This shortcoming reflects the limited counting ability of the generation model, as observed in prior research and experiments \cite{Ref018}.

To overcome this problem, we added a critic into the image reconstruction process to ensure reliable performance. GPT-4V has strong image analysis capabilities for tasks such as object counting and detection. During the critique process, GPT-4V will initially provide a ``YES" or ``NO" response regarding the accuracy of the types and quantities of all objects in the generated image. Subsequently, GPT-4V generates a detailed prompt to guide DALL-E-3 in producing a more accurate reconstruction. For example, given the example in Fig. \ref{fig:figure5}a, GPT-4V will give a ``NO" at the beginning of the response and then tell the image generation model (i.e., Dall-E-3) that the number of orange of should be three instead of four. With the modification pointed out by GPT-4V, DALL-E-3 starts a new round of image generation. However, DALL-E-3 cannot always guarantee that the modified image after this iteration is correct. Therefore, the process will continue to iterate until GPT-4V outputs ``YES" or the number of iterations reaches an iteration upper bound.


\begin{table*}[htbp]\label{table:1}
\renewcommand{\arraystretch}{1.25}
\captionsetup{font={small}}
\caption{Performance comparisons for different models/scheme over $D$ and $D_{nOOD}$.}
\center
\begin{tabular}{p{4.25cm}<{\centering}
|p{1.0cm}<{\centering}p{1.0cm}<{\centering}p{1.25cm}<{\centering}
|p{1.0cm}<{\centering}p{1.0cm}<{\centering}p{1.25cm}<{\centering}
}\hline
\multicolumn{1}{c|}{\multirow{2}{*}{Tested Model/Scheme}} & \multicolumn{3}{c|}{Performance on $D_{nOOD}$} & \multicolumn{3}{c}{Performance on $D$} \\ \cline{2-7} 
 & Precision  & Recall & F1 Score & Precision  & Recall & F1 Score \\ \hline
\multirow{1}{*}{YOLOv8(expert AI solely)}& 0.6095 & 0.6177 & 0.6136 & 0.5176 & 0.5670 & 0.5412   \\ 
\multirow{1}{*}{InstructBLIP (general AI solely)}& 0.7420 & 0.5252 & 0.6150 & 0.7333 & 0.5620 & 0.6363 \\ 
\multirow{1}{*}{Plan A – Plan B}& 0.7727 & 0.6869 & 0.7272 & 0.7468 & 0.6903 & 0.7174 \\ \hline

\end{tabular}
\label{tab:table1}
\end{table*}

\section{Experiments and Validations}\label{sec-V}
\subsection{Evaluations of the Plan A -- Plan B framework}
This subsection evaluates the anti-OOD performance of the proposed Plan A -- Plan B framework. We crop the images in the original COCO validation dataset based on the ground truth annotations.\footnote{This is because the testing set of the COCO dataset does not provide ground truth annotations. Instead, the dataset asks users to submit their identification results to a cloud serve for evaluation. In our experiment, we need to leverage the ground truth within an image to construct the experimental dataset. Hence, we choose the validation set. Note that the validation set was never used in our model training, i.e., no data leakage happens in our training process.} And then we remove images that are too small to be identified even for human eyes. After the above operations, we obtain a new dataset $D$ containing a total of 14,423 images, which has 1,183 OOD images (i.e., animal classes). Let us denote the set of OOD and non-OOD images by $D_{OOD}$ and $D_{nOOD}$, respectively. To comprehensively evaluate the framework, we compared the object recognition accuracy of YOLOv8 (Plan A), InstructBLIP (Plan B), and Plan A -- Plan B scheme on dataset $D$ and $D_{nOOD}$, respectively. We did not test pure OOD cases, as this may not be very practical (more or less, an actual scenario should contain a few familiar cases).

Table \ref{tab:table1} presents the precision, recall, and F1 score of different models in $D$ and $D_{nOOD}$. For the non-OOD dataset, YOLOv8 can achieve a precision of 60.95\% and a recall of 61.77\%. Its lower precision indicates that it overfits on the identification of high-proportion classes, thus leading to misidentification and reduced accuracy. InstructBLIP achieves 74.2\% and 52.52\% for precision and recall in the same tested dataset. That shows InstructBLIP is more ``cautious" than YOLOv8 in identifying samples of high-proportion classes, i.e., it prefers to sacrifice recall to ensure precision. For the model's general performance (i.e., F1 score), YOLOv8 and InstructBLIP have similar performance, reaching 0.6136 and 0.6150, respectively. The Plan A -- Plan B scheme scheme combines two models'  advantages effectively and achieves the best performance in all aspects (i.e., both precision, recall, and F1 score).

For the comprehensive dataset $D$, the performance of YOLOv8 drops obviously due to the involvement of OOD cases. InstructBLIP, on the other hand, demonstrates less obvious changes due to its anti-OOD ability. The Plan A-Plan B scheme shows higher performance than the above two single-model schemes, reaching the highest performance in terms of precision, recall, and F1 score. This experimental observation indicates the superiority of our Plan A-Plan B scheme in the object identification process within semantic encoding.

\begin{figure}[htbp]
\centering
\includegraphics[width=0.32\textwidth]{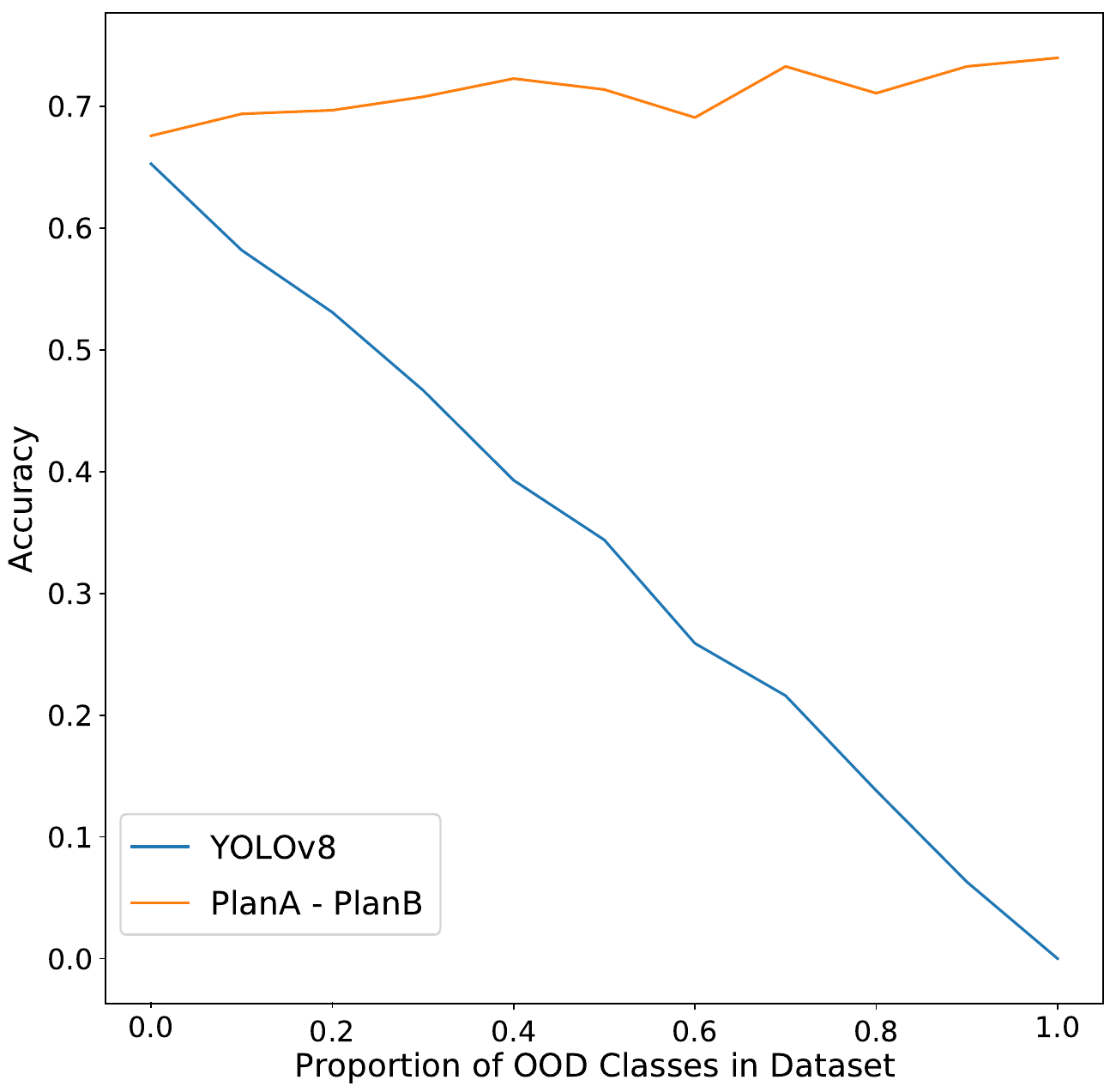}
\captionsetup{font={small}}
\caption{The accuracies of YOLOv8 and the Plan A -- Plan B scheme under different proportions of OOD cases in the dataset.}
\label{fig:figure6}
\end{figure}

To further evaluate the anti-OOD performance of different models/schemes, we create a group of new testing data by selecting a total of 100 images from $D_{nOOD}$ and $D_{OOD}$ with different proportions. We adjust the proportion of OOD cases in the tested dataset and compare the recognition accuracy of YOLOv8 and the Plan A -- Plan B scheme. Fig. \ref{fig:figure6} shows that YOLOv8's performance continues to decrease as the proportion of OOD cases increases. In contrast, the Plan A -- Plan B scheme can maintain relatively stable performance regardless of the proportion of OOD cases in the tested data. This further justifies the strong anti-OOD performance of the proposed framework.

\subsection{Experiments of the Bayesian optimization scheme}
This section discusses how we select the best-fit embedding network for our Bayesian optimization scheme and presents our evaluation of the scheme's performance. In general, the performance of our scheme should be evaluated from two angles: to what degree can our scheme correct those unsuccessful identifications made by InstructBLIP? Will the scheme miscorrect cases that were successfully identified by InstructBLIP? To answer these questions, we built the following dataset for testing. To begin with, we use the YOLO model to analyze all images within the COCO validation dataset and record the model's confidence over each object. We treat objects with confidence level $\rho$ smaller than 0.7 as unconfident cases, while the rest are confident cases. We focus on those unconfident cases, as we do not rely on InstructBLIP (and the Bayesian scheme therein) on those confident ones. To test the Bayesian scheme, we want images with at least one unconfident object as the optimization target and at least one confident object as the background knowledge. After the above filtering process, we crop the selected images according to the ground truth for InstructBLIP's later processing. Finally, after the processing of InstructBLIP, we obtain a dataset for testing the Bayesian optimization scheme, which contains 1) 572 images that are incorrectly identified by InstructBLIP, and 2) 444 images that are successfully ``rescued” by InstructBLIP without Bayesian optimization.

For the rest of the discussion, let us denote the probability that InstructBLIP revises the wrong identification to a correct one after Bayesian optimization by $R_+$. Furthermore, we denote the probability of turning a correctly identified case into an erroneous one after Bayesian optimization by $R_-$.

\begin{figure}[htbp]
\centering
\includegraphics[width=0.45\textwidth]{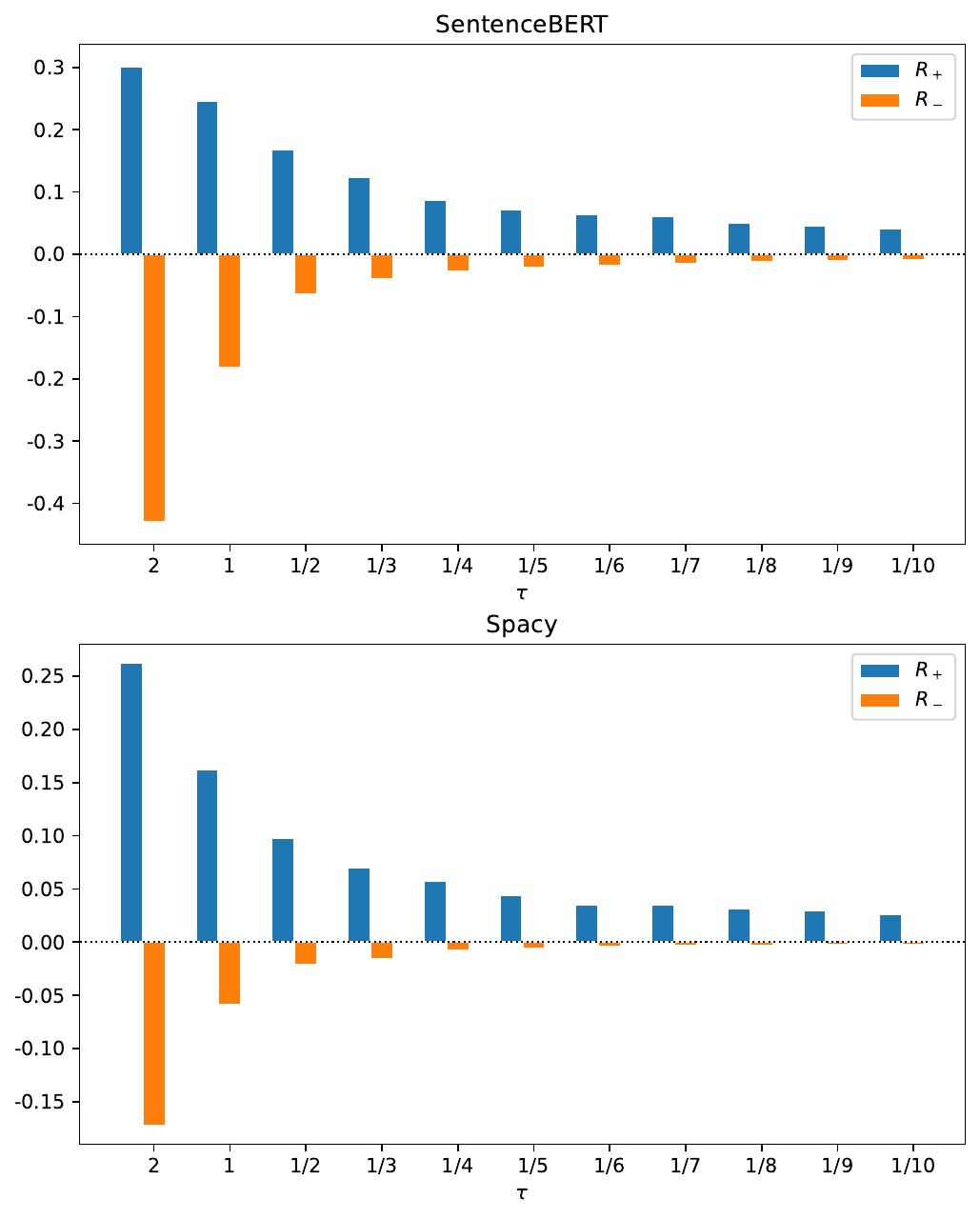}
\captionsetup{font={small}}
\caption{The correction rate of different embedding networks versus index $\tau$. Note that we plot $-R_-$ here for better visual presentation.}
\label{fig:figure7}
\end{figure}

Fig. \ref{fig:figure7} presents experimental results of $R_+$ and $R_-$ when two example embedding networks are tested on the dataset described above with different $\tau$ values considered. An interesting observation we see from the figure is that both $R_+$ and $R_-$ decrease if we set $\tau$ lower, which indicates that the Bayesian scheme is conservative in revising the original identification outputs. That results in minor corrections and limited gains. For example, if we set $\tau = 1/10$ for Spacy, although the scheme has a good performance in terms of $R_-$ ($R_-=0$, i.e., no mistaken overturning happens), it makes very limited contribution in terms of $R_+$. When $\tau$ increases, on the other hand, the scheme becomes aggressive in modifying the original identification output extracted with the CET scheme. For example, if we set $\tau = 2$ for SentenceBERT, our scheme becomes very reckless in overturning the previous conclusion, resulting in a $R_-$ that is even larger than $R_+$.

The above experimental results lead us to an interesting discussion regarding the performance evaluation of an NLP embedding network for our task. Ideally, we want higher $R_+$ and lower $R_-$ simultaneously. Yet, given a fixed embedding network, we see from the above experiments that we have to make a trade-off between higher $R_+$ and lower $R_-$ by adjusting $\tau$. In light of this, when answering the question ``which embedding network is the best fit for our task", we have to consider both sides of the coin. For example, we cannot simply conclude that one embedding network is more suitable than another because it has higher $R_+$, as it may come with the price of higher $R_-$ (and vice versa).

\begin{figure}[htbp]
\centering
\includegraphics[width=0.32\textwidth]{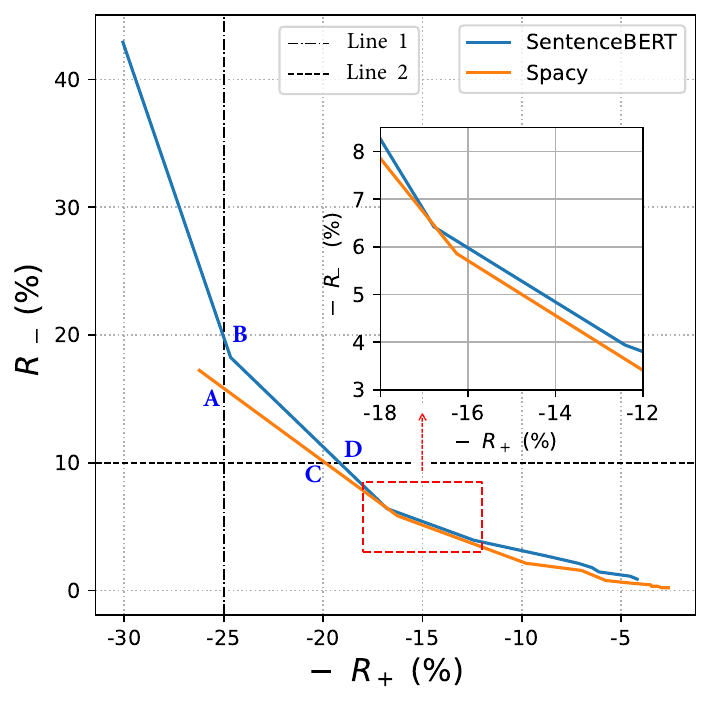}
\captionsetup{font={small}}
\caption{A comparison of SentenceBERT's and Spacy's Pareto efficiency.}
\label{fig:figure10}
\end{figure}

A more systematic way to compare the two embedding networks is to look at their ``Pareto Performance" that considers both R+ and R- at the same time \cite{Ref044c, Ref044a, Ref044b}. Given an embedding network, we can obtain a series of $\left( R_+ , R_- \right)$ pairs by changing the setting of $\tau$. With these $\left( R_+ , R_- \right)$ pairs, we plot the performance curve of the tested embedding networks in Fig. \ref{fig:figure10}, where the y-axis is $R_-$, and the x-axis is $-R_+$.\footnote{Similar to Fig. \ref{fig:figure7}, where we use $-R_-$ in plotting the figure for better illustration, Fig. \ref{fig:figure10} uses $-R_+$ as the x-axis for clearer visual presentation and discussion associated with it.}

In Fig. \ref{fig:figure10}, we first assume a constant $R_+ = 0.25$ as in Line One. From Point A and Point B in the figure, we see that Spacy has lower $R_-$ (which is more desired) than SentenceBERT when they have the same $R_+$ performance. In that situation, we say Spacy is better than SentenceBERT under a constant $R_+ = 0.25$ setting. Furthermore, if we assume a constant $R_- = 0.10$ as in Line Two, from Point C and Point D in the figure, we see that Spacy has a higher $R_+$ (which is more desired) than SentenceBERT when they have the same $R_-$ performance. In that situation, we say Spacy is better than SentenceBERT under a constant $R_- = 0.10$ setting.

More generally, if the performance curve of embedding network A is consistently below that of embedding network B, then we say that A is a better fit for our task than B, as we can always find network A with a lower $R_-$ when both A and B has the same $R_+$. With the above analysis, we see that Spacy is obviously a better fit for our Bayesian optimization scheme, given that its performance curve is always below that of SentenceBERT.

The above benchmarking process compares the Pareto performance of the two networks. After selecting Spacy as the embedding network, we now take a closer look at the performance of our Bayesian optimization scheme on the tested dataset. Let us define the performance metric of our Bayesian optimization scheme as,
\begin{equation}
R = - \varepsilon_+ R_+ + \varepsilon_- R_- 
\label{def_R}
\end{equation}
where $\varepsilon_+$ and $\varepsilon_-$ are the proportion of images that are incorrectly and correctly identified by the original model before applying InstructBLIP with Bayesian optimization, respectively. For our tested dataset described at the beginning of this subsection, 572 out of 1016 images are incorrectly identified, while the rest 444 images are successfully ``rescued" by InstructBLIP. Hence, we have $\varepsilon_+ = 572/1016 = 0.5630$ and $\varepsilon_- = 444/1016 = 0.4370$.

With the definition of $R$ in (\ref{def_R}), we see that the original multi-objective optimization problem (i.e., higher $R_+$ and lower $R_-$) is equivalent to the minimization of $R$, which is an easier single-objective optimization problem. For a visual illustration of how we find the optimal $R$, we rewrite (\ref{def_R}) as
\begin{equation}
R_- = (-\frac{\varepsilon_+}{\varepsilon_-}) \cdot (-R_+) + \frac{R}{\varepsilon_-}
\label{rewrite_eq}
\end{equation}

\begin{figure}[htbp]
\centering
\includegraphics[width=0.34\textwidth]{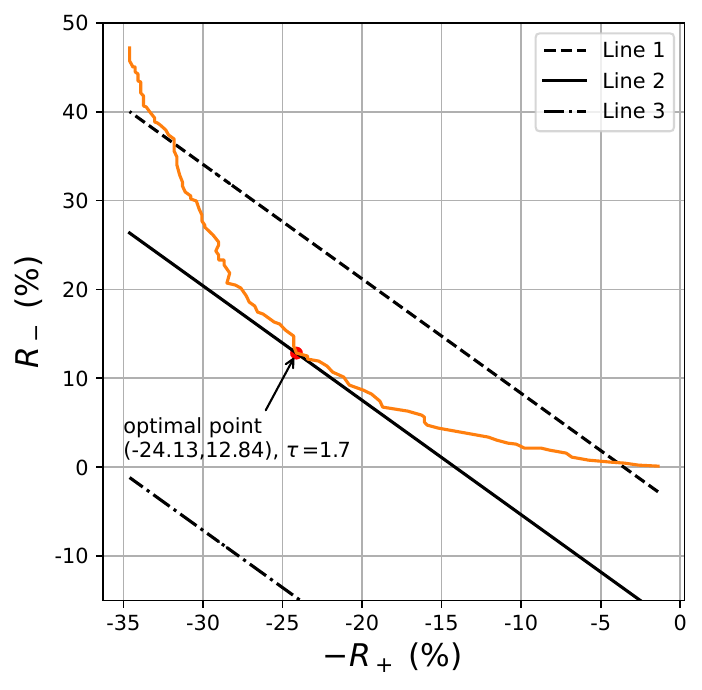}
\captionsetup{font={small}}
\caption{The Pareto curve of our Bayesian optimization scheme (using Spacy as the selected network). Here the Pareto curve is obtained with a $\tau$ ranges from 0 to 5, using a testing step of 0.025.}
\label{fig:figure11}
\end{figure}

In Fig. \ref{fig:figure11}, where the y-axis is $R_-$ and the x-axis is $-R_+$, we plot 1) the Pareto curve of our scheme with Spacy and 2) straight lines obtained according to (\ref{rewrite_eq}) with different $R$. The setting of $R$ is infeasible for our Bayesian optimization scheme if the corresponding straight line (e.g., Line 3) does not touch the Pareto curve. Among those feasible lines (e.g., Line 1 and Line 2), we are interested in the one that corresponds the minimum $R$.

The tangent line of the Pareto curve with slope $-{\varepsilon_+}/{\varepsilon_-}$, i.e., Line 2, results in the smallest $R$. That is, by moving from Line 1 to Line 2, we are gradually reducing the y-intersection of the line, which means smaller $R$ according to (\ref{rewrite_eq}). Finally, when the line touches the Pareto curve at only one point, we get the optimal $R$. Further lower the line makes $R$ infeasible.

In our tested dataset, when we set $\tau$ as 1.7, the corresponding line touches the curve at the only point $\left(-24.13,12.84\right)$. Substituting the point into (\ref{def_R}), we have $R = -7.98\%$, which is the optimal $R$ we are looking for.

\subsection{Evaluations of the generate-criticize framework}
This section evaluates the performance of the generate-criticize framework. We are interested in how the iteration limit of the image critic affects the image generation process. We first randomly select an object from the 80 possible classes and randomly generate quantity $N$ that describes the number of this object.\footnote{Here we assume $N\geq 3$, given that our prior observations suggest that a text-to-image model tends to generate incorrect object counts when the number of objects is larger than three.} We construct a prompt based on the selected object and the corresponding $N$. We then leverage the generate-criticize framework to build the corresponding image. After the image is constructed, we manually determine whether the generated images were consistent with the semantics in the prompt. For each iteration limit, we conduct the above process 100 times to calculate the average accuracy.

\begin{figure}[htbp]
\centering
\includegraphics[width=0.4\textwidth]{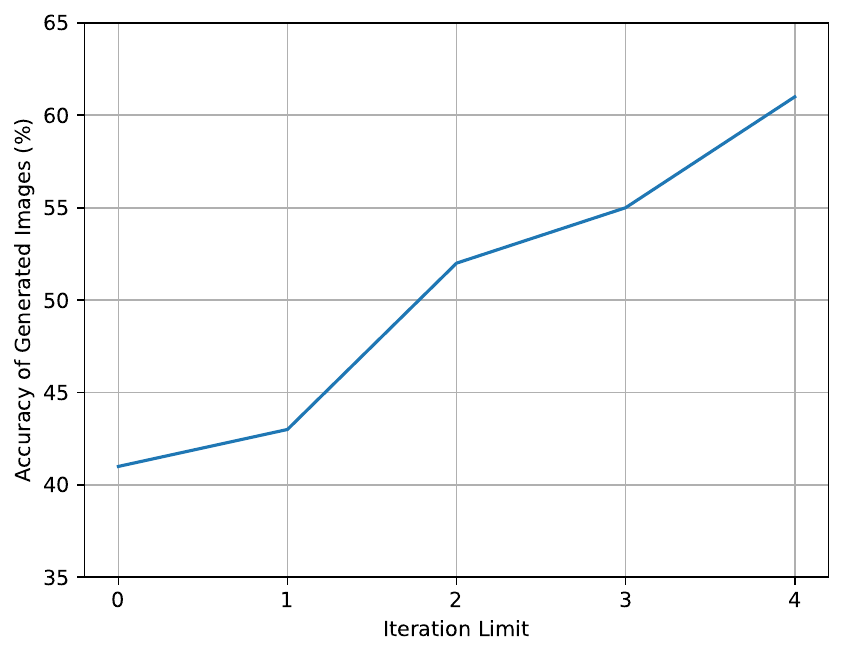}
\captionsetup{font={small}}
\caption{The accuracy of generated images corresponding to different iteration limits.}
\label{fig:figure8}
\end{figure}

Fig. \ref{fig:figure8} shows that as the iteration limit of the image critic increases, the accuracy of the generated images continues to improve. When the iteration limit is set to 4, the corresponding accuracy rate is nearly 20\% higher than the scheme without the critic. These results indicate that the generate-criticize framework can significantly improve the reliability of image reconstruction by leveraging the image critic to provide feedback and guide the generation process.

\subsection{The system's anti-noise performance}
Finally, Subsection D studies the system's anti-noise performance by investigating the semantic loss (see (\ref{loss_function}) for the definition of the system's semantic loss) under different channel noise. Our detailed experimental setup is as follows: in the training process of the T2T JSCC encoder and decoder, we assume an additive white Gaussian noise (AWGN) channel with an SNR of 10dB. While in the testing process, we consider noise conditions that are significantly larger than the expected one (with the SNR ranging from -20dB to 5dB).

\begin{figure}[htbp]
\centering
\includegraphics[width=0.4\textwidth]{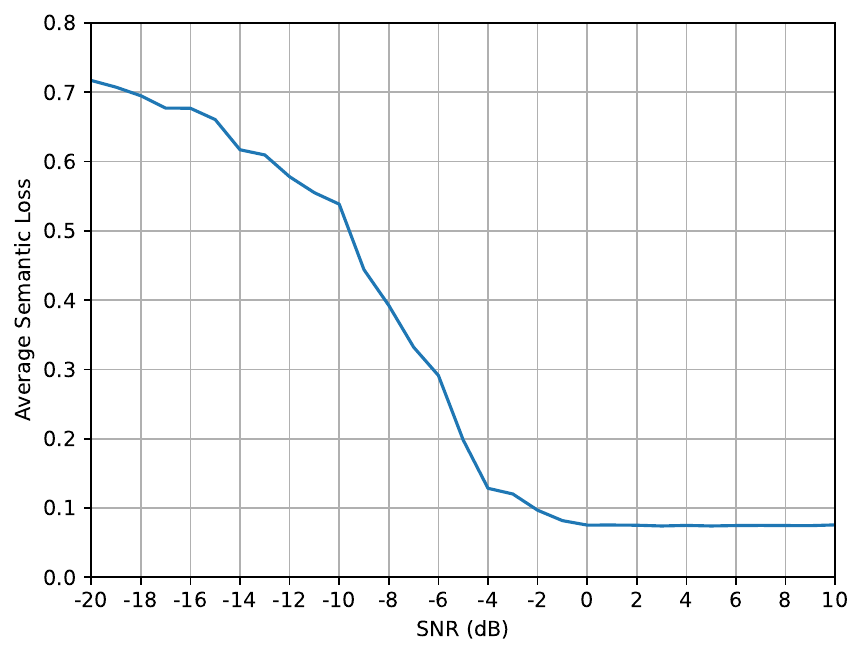}
\captionsetup{font={small}}
\caption{The average loss performance of the whole system when transmitting 100 images (including 8 OOD cases) under different noise conditions.}
\label{fig:figure9}
\end{figure}

In the experiment, we assume $\alpha=0.1$ and transmit images through the tested AWGN channel. Figure \ref{fig:figure9} illustrates the average semantic loss when different channel conditions are considered. As we can see from the figure, even if the SNR is as low as -2dB, which is far from the SNR setting in the training process, the average loss remains very close to that of 10dB-SNR case. This experimental result demonstrates the ultra-robustness of the proposed scheme in the face of the strong channel noise.

\section{Conclusion}\label{sec-VI}
This paper explored the application of MLLM in semantic communication to address the OOD problem in the semantic compression process. Our proposed ``Plan A - Plan B" framework leverages the broad knowledge scope and strong generalization ability of MLLMs to assist conventional ML models in semantic encoding when encountering OOD cases. Furthermore, we introduced a Bayesian optimization scheme that reshapes the probability distribution of the MLLM's inference process based on the image's contextual information, significantly enhancing the MLLM's performance in accurate semantic compression. On the receiver side, our ``generate-criticize" framework utilizes the cooperation of multiple MLLMs to improve the reliability of image reconstruction. Experimental results demonstrate the effectiveness of our proposed frameworks in terms of the accuracy of the semantic compression process, the probability of correct image generation, and the robustness in the face of strong channel noise. These findings highlight the potential of MLLMs in enabling reliable and efficient semantic communication for next-generation wireless networks. Future work could explore the application of our proposed frameworks to other types of data, such as video and audio, with the assistance of more powerful MLLMs. Additionally, the development of more advanced JSCC schemes, which could further enhance the practicality of MLLM-empowered semantic communication systems, might also be an interesting topic for the community.

\bibliographystyle{IEEEtran}
\bibliography{Main}

\end{document}